\documentclass[a4paper,10pt,dvipsnames]{article}
\pdfoutput=1 
\usepackage{jcappub}
\usepackage[T1]{fontenc}
\usepackage{appendix}
\usepackage{enumitem}
\usepackage{booktabs}
\usepackage{multirow}
\usepackage{siunitx}
\usepackage{orcidlink}
\usepackage{arydshln}
\usepackage{caption}
\usepackage{subfig}
\usepackage{graphicx}
\usepackage{bm}
\usepackage[normalem]{ulem}
\usepackage{float}

\newcommand{\agtext}[1]{\textcolor{cyan}{#1}}
\newcommand{\agcomm}[1]{\textcolor{Bittersweet}{[AG: \textit{#1}]}}

\newcommand{\dgw}{d_L^{\rm GW}}
\newcommand{\dem}{d_L^{\rm EM}}
\newcommand{\xgw}{x_{\rm GW}}
\newcommand{\xem}{x_{\rm EM}}
\newcommand{\xz}{x_{\rm z}}
\newcommand{\dx}{d_x}
\newcommand{\dy}{d_y}
\newcommand{\dz}{d_z}
\newcommand{\rc}{R_c}
\newcommand{\al}{\alpha}
\newcommand{\blmb}{\bm{\lambda}}
\newcommand{\bxi}{\bm{\xi}}
\newcommand{\bxic}{\bm{\xi^c}}

\title{\boldmath Constraining cosmological extra dimensions with gravitational wave standard sirens: from theory to current and future multi-messenger observations}

\author[a,b]{Maxence Corman\orcidlink{0000-0003-2855-1149},}
\author[c]{Abhirup~Ghosh\orcidlink{0000-0002-2112-8578},}
\author[d]{Celia~Escamilla-Rivera\orcidlink{0000-0002-8929-250X},}
\author[e]{Martin A.~Hendry\orcidlink{0000-0001-8322-5405},}
\author[f]{Sylvain Marsat\orcidlink{0000-0001-9449-1071},}
\author[g]{and Nicola Tamanini\orcidlink{0000-0001-8760-5421}}

\affiliation[a]{Perimeter Institute for Theoretical Physics, Waterloo, ON N2L 2Y5, Canada.}
\affiliation[b]{Department of Physics \& Astronomy, University of Waterloo, Waterloo, ON N2L 3G1, Canada}
\affiliation[c]{Max Planck Institute for Gravitational Physics (Albert Einstein Institute), Am M\"uhlenberg 1, Potsdam 14476, Germany}
\affiliation[d]{Instituto de Ciencias Nucleares, Universidad Nacional Aut\'onoma de M\'exico, Circuito Exterior C.U., A.P. 70-543, M\'exico D.F. 04510, M\'exico.}
\affiliation[e]{SUPA, School of Physics and Astronomy, University of Glasgow, Glasgow G12 8QQ, United Kingdom.}
\affiliation[f]{Universit\'{e} de Paris, CNRS, Astroparticule et Cosmologie, F-75013 Paris, France}
\affiliation[g]{Laboratoire des 2 Infinis - Toulouse (L2IT-IN2P3), Universit\'e de Toulouse, CNRS, UPS, F-31062 Toulouse Cedex 9, France}
\emailAdd{mcorman@perimeterinstitute.ca}
\emailAdd{abhirup.ghosh@aei.mpg.de}
\emailAdd{celia.escamilla@nucleares.unam.mx}
\emailAdd{martin.hendry@glasgow.ac.uk}
\emailAdd{marsat@apc.in2p3.fr}
\emailAdd{nicola.tamanini@l2it.in2p3.fr}

\abstract
{The propagation of gravitational waves (GWs) at cosmological distances offers a new way to test the gravitational interaction at the largest scales.
Many modified theories of gravity, usually introduced to explain the observed acceleration of the universe, can be probed in an alternative and complementary manner with respect to standard electromagnetic (EM) observations.
In this paper we consider a homogeneous and isotropic cosmology with extra spatial dimensions at large scales, which represents a simple phenomenological prototype for extra-dimensional modified gravity cosmological models.
By assuming that gravity propagates through the higher-dimensional spacetime, while photons are constrained to the usual four dimensions of general relativity, we derive from first principles the relation between the luminosity distance measured by GW detectors and the one inferred by EM observations.
We then use this relation to constrain the number of cosmological extra dimensions with the binary neutron star event GW170817 and the binary black hole merger GW190521.
We further provide forecasts for the Laser Interferometer Space Antenna (LISA) by simulating multi-messenger observations of massive black hole binary (MBHB) mergers.
This paper extends and updates previous analyses which crucially neglected an additional redshift dependency in the GW-EM luminosity distance relation which affects results obtained from multi-messenger GW events at high redshift, in particular constraints expected from LISA MBHBs.}

\begin{document}
\maketitle
\flushbottom


\section{Introduction}

The first observation of Gravitational Waves (GWs) from binary black-hole coalescences \cite{PhysRevLett.116.061102}, the first observation of a neutron star binary coalescence \cite{TheLIGOScientific:2017qsa}, and the identification of an explicit electromagnetic (EM) counterpart \cite{GBM:2017lvd} have opened a new era of GW and multi-messenger astronomy.
In the near future, with advanced LIGO and advanced Virgo reaching their design sensitivity \cite{2018LRR....21....3A,LIGOScientific:2021hoh,LIGOScientific:2021tsm}, and KAGRA\footnote{\url{https://gwcenter.icrr.u-tokyo.ac.jp/en/}~\cite{Aso:2013eba}} and LIGO-India\footnote{\url{https://www.ligo-india.in} (\url{https://dcc.ligo.org/LIGO-M1100296/public})} joining the global network of second-generation ground-based detectors, we expect GW detections to take place on an almost daily basis.
Furthermore, in the 2030s the space-based interferometer LISA\footnote{\url{https://lisa.nasa.gov}} and third-generation (3G)\footnote{\url{https://gwic.ligo.org/3Gsubcomm/}} ground-based interferometers such as the Einstein Telescope (ET)\footnote{\url{https://www.einsteintelescope.nl}} and Cosmic Explorer (CE)\footnote{\url{https://cosmicexplorer.org}}, will be capable of detecting large numbers of coalescing compact binaries at cosmological redshifts, $z \gg 1$.
These developments will open up excellent opportunities for constraining cosmological parameters and testing cosmological models beyond the standard framework of General Relativity.

The detection of GWs from a coalescing binary allows for a direct measurement of its luminosity distance $\dgw$, just as the observation of an EM counterpart leads to a direct measurement of the source redshift. Assuming a specific cosmological model of the Universe, information on the redshift is used to infer the EM luminosity distance $\dem$. A comparison of the observed (GW) and inferred (EM) luminosity distances can then provide us with constraints on the parameters of our cosmological model. Analogous to standard candles, these sources are referred to as \textit{standard sirens} \cite{Holz_2005,Zhang:2019ylr}, and the first results constraining the Hubble constant using GW sirens were reported in~\cite{Abbott:2017xzu}. However, the observation of an EM counterpart (a so-called `bright siren') is rare. Therefore, in recent times there has been considerable work focussing on measurements of cosmological parameters using `dark sirens', where the redshift of the host galaxy is inferred statistically from galaxy surveys or assumptions about the population~\cite{Abbott_2021H0}.

In the $\Lambda$CDM standard model of cosmology, the late time cosmic acceleration is explained by a non-zero but very small cosmological constant that opposes the self-attraction of pressure-less matter and causes this accelerated expansion. Notwithstanding the broad success of the $\Lambda$CDM paradigm, it leaves several significant unresolved tensions between the values of certain cosmological parameters (not least the Hubble constant itself) inferred from different data sets and cosmological probes \cite{DiValentino:2021izs,DiValentino:2020zio}.
The adoption of $\Lambda$CDM fundamentally demands the inclusion of cold dark matter (CDM) and a cosmological constant ($\Lambda$). However the physical origin of these two largest contributions to the energy content of the late-time universe remains a mystery. Therefore it is both highly relevant and timely to study alternative explanations for the cosmic late-time acceleration. In that regard, several proposals have been investigated -- including exotic fluids \cite{DiValentino:2020evt}, modified gravity \cite{Clifton:2011jh}, extended gravity theories \cite{Bahamonde:2021gfp} and higher dimensional theories \cite{Padilla:2004uj,Dvali:2000hr,Fang:2008kc}. In the last category of high-dimensional theories, we assume standard matter to be confined to a 3D spatial brane, 
while gravity propagates in all dimensions. This leads to a predicted deviation from GR gravity at large length scales, and the proposal of a possible \textit{gravitational leakage} at cosmological distances, providing an additional damping of the GW amplitude as it propagates in the higher-dimensional universe.
A comparison between the luminosity distances measured by GW detectors and inferred from EM observations, respectively, allows us to constrain this analogous damping of the GWs, and consequently to probe the gravitational leakage.
Such an effect is similar to what other modified gravity models predict, which can thus be tested in exactly the same way; see e.g.~\cite{Belgacem:2018lbp,eLISA2,Calcagni:2019kzo,Calcagni:2019ngc,Mastrogiovanni:2020gua,Mukherjee:2020mha,Mastrogiovanni:2020mvm}.
Some modifications to the signal’s attenuation with luminosity distance due to higher-dimensions have also been studied in \cite{Pardo:2018ipy}, where -- given a prior on $H_0$ -- constraints on $D$, the number of spacetime dimensions, were derived using the observations of GW170817 in both GWs and EM. The results found that $D = 4.02^{+0.07}_{-0.10}$, using the SH0ES prior for $H_0$, and $D\approx 3.98^{+0.07}_{-0.09}$,  using the Planck prior -- in both cases 68\% credible regions.
A further analysis by the LIGO and Virgo collaborations found consistent results \cite{Abbott:2018lct}. Given the proximity of GW170817, however, these constraints on $D$ apply only at very low redshift, $z < 0.01$.
Note moreover that at low redshift one can also use another complementary way to test higher-dimension models, by counting the number of GW detections and looking at their distribution in SNR \cite{Calabrese:2016bnu,Garcia-Bellido:2016zmj}. This method has the advantage of not requiring any EM counterpart; however it requires a large number of low-redshift GW sources and it assumes a specific spatial distribution for them.

The effects of gravitational leakage, and the feasibility of using it to place constraints on higher-dimensional models, were further studied in~\cite{Corman:2020pyr}, in the context of future observations of high-redshift sirens by LISA -- thus exploiting the capability of LISA to probe modified gravity on large cosmological scales. This paper considered the same phenomenological models explored in \cite{Pardo:2018ipy} and also the particular case of the Dvali, Gabadadze and Porrati (DGP) model. It was found that the extent to which LISA will be able to place limits on the number of spacetime dimensions and other cosmological parameters characterising modified gravity will strongly depend on the actual number and redshift distribution of sources, together with the uncertainty on the GW measurements. In the most optimistic scenarios, however, it was found that LISA has the potential to constrain the number of spacetime dimensions to about 1\% and the scale beyond which gravity is modified to better than about 10\%.

In this paper we study the consequences of higher dimensional theories with non-compact extra dimensions, re-deriving the phenomenological implications from first theoretical principles, re-visiting the constraints that may be obtained from observations of GW170817 and forecasting the capability of LISA to constrain these theories from future high-redshift observations GW standard sirens with an associated EM counterpart. 
This paper is organised as follows: 
in Sec.~\ref{sec:theory} we introduce the equations for GW propagating in higher-dimensions and derive the relation between the luminosity distance of a cosmological source as measured by GW and EM observations, showing that a new additional redshift dependence, not considered in previous literature, must be taken into account. 
In Sec.~\ref{sec:bayesian} we outline the Bayesian representation we use to develop the inference method in our models.
In Sec.~\ref{sec:constraints} we apply the Bayesian approach taking as input the LIGO-Virgo measurement of GW170817, updating the current constraints derived from this event.
In Sec.~\ref{sec:LISA} we present new forecasts for LISA by applying the Bayesian approach to MBHB events, and by showing how the new redshift factor in the luminosity distance relation affects expected results at high redshift.
Our main conclusions are then presented and summarised in Sec.~\ref{sec:conclusion}.


\section{Gravitational wave propagation in a higher-dimensional universe: derivation of the luminosity distance relation}
\label{sec:theory}

In this section we discuss how GWs propagate in a $D$-dimensional universe, with $D=N+1$ representing the number of spacetime dimensions.
By explicitly solving the GW propagation equation in a $D$-dimensional cosmological universe, we derive the dependency of the GW amplitude over the observed luminosity distance and its relation with the 4-dimensional luminosity distance measured by EM observations.
In what follows we show that the relation between the GW and EM luminosity distances, has an additional redshift dependency which has so far been neglected in the literature.
As we show in the next sections of the paper, taking into account this new redshift factor is fundamental to obtain correct constraints on the actual number of spacetime dimensions, especially at cosmological distances.

Let us start by considering a binary system emitting GWs in a standard $D=4$ Minkowski spacetime.
At the lowest (Newtonian) order, far away from the source (the so-called wave-zone \cite{Maggiore:2007ulw}) the gravitational waveform is given by:
\begin{equation}
	h_\times (t_s) = \frac{4}{r_3} \left( G \mathcal{M}_c \right)^{5/3} \left( \pi f_s \right)^{2/3} \cos\theta \sin\Phi(t_s) \,,
	\label{eq:wf_s}
\end{equation}
where $h_\times$ is the $\times$(cross)-polarization of the GW. There exists a similar expression for the $+$(plus)-polarisation (for simplicity, here we only consider one polarisation mode).
In Eq.~\eqref{eq:wf_s} $r_3$ is the 3-dimensional radial coordinate distance\footnote{Assuming spherical coordinates centered at the source for the spatial brane of the full 4-dimensional spacetime.} and the other quantities have their usual meaning: $t_s$, $f_s$, $\mathcal{M}_c$, $\theta$ and $\Phi$ denote the time and frequency at the source, the chirp mass, the inclination angle and the GW phase, respectively.
Eq.~\eqref{eq:wf_s} represents the standard GR waveform describing GWs as seen by an observer at non-cosmological distances.
In what follows we will always assume that at non-cosmological distances (say $\lesssim$ 100 kpc) the universe is 4-dimensional, and that GW generation and propagation are well described by the usual expressions derived in GR. 
This assumption is necessary in order to satisfy constraints from current tests of GR, such as Solar System experiments for example \cite{Liu:2000zq,Overduin:2000gr}.
We will consider higher-dimensions only at cosmological distances, implicitly assuming that a transition from 4 to $D$ dimensions should happen at some specific scale.
In what follows we need thus to take into account cosmological and higher-dimensional effects simultaneously.

Let us first review how the waveform \eqref{eq:wf_s} is observed at cosmological distances.
In standard 4 dimensions, to account for cosmology, i.e.~for the expansion of the universe, one must simply replace $M_c$ and $r_3$ respectively with $\mathcal{M}_{cz}$ and $d_L^{\rm (4)}$, the so-called redshifted chirp mass and the standard 4-dimensional luminosity distance, given by
\begin{equation}
	d_L^{(4)} = a_0 (1+z) r_3 \,,
	\label{eq:def_dL4}
\end{equation}
where $a_0$ is the value of the scale factor today (usually set equal to one) and $z$ is the cosmological redshift.
Moreover one must now take into account that time and frequency at the observer, namely $t_o$ and $f_o$, are different from time and frequency at the source, $t_s$ and $f_s$, due to the cosmological redshift.
All this implies that at cosmological distances, in a 4-dimensional universe, GWs are characterised by the lowest-order waveform (see e.g.~\cite{Maggiore:2007ulw})
\begin{equation}
	h_\times (t_o) = \frac{4}{d_L^{(4)}} \left( G \mathcal{M}_{cz} \right)^{5/3} \left( \pi f_o \right)^{2/3} \cos\theta \sin\Phi(t_o) \,.
	\label{eq:wf_o}
\end{equation}
We now want to understand how this expression reads in a general $D$-dimensional universe.

Let us start again by assuming that far away from the source the spacetime becomes effectively $D$-dimensional, where $D=N+1$ is a number greater than 4.
The (gravitational) wave equation in this spacetime can again be derived from the Einstein field equations, yielding simply
\begin{equation}
	\Box h_{\mu\nu} = 0 \,,
\end{equation}
where we assumed to be in vacuum and imposed the usual Lorenz gauge.
Note however that the Greek indices now run from 0 up to $D-1$,
as we are considering a $D$-dimensional spacetime.
Taking the standard geometric optics approximation, the GW solution of the equation above can be written as (see e.g.~\cite{Laguna:2009re})
\begin{equation}
	h_{\mu\nu} = e_{\mu\nu} \mathcal{A} e^{i\Phi / \epsilon} \,,
\end{equation}
where $e_{\mu\nu}$ is a polarization tensor, $\mathcal{A}$ is the amplitude of the wave and $\Phi$ is again the phase.
Here $\epsilon$ is a small parameter over which the equations are to be expanded in the geometric optics domain.
At the leading orders in $\epsilon$ one gets the following equations
\begin{align}
	k_\mu k^\mu = 0 \,, \quad\mbox{and}\quad \nabla_\mu \left( \mathcal{A}^2 k^\mu \right) = 0\,,
	\label{eq:001}
\end{align}
where $\nabla$ is the covariant derivative of the $D$-dimensional spacetime.
The first one of these relations is just saying that GWs propagate along null geodesics of the higher-dimensional spaetime, while the second one provides a conservation equation for the (square) amplitude along the geodesic.
We can now solve Eqs.~\eqref{eq:001} in a $D=N+1$ Minkowski spacetime with invariant element
\begin{equation}
	ds^2 = -dt^2 + \sum_{i = 1}^{N} dx_i^2 \,.
	\label{eq:le_Minkowski_D}
\end{equation}
Integrating the second of Eqs.~\eqref{eq:001} one gets (see Appendix~\ref{app:GW_propag_D})
\begin{equation}
	\mathcal{A} \propto r_N^{-(D-2)/2} \,,
	\label{eq:003}
\end{equation}
where $r_N^2 = \sum_{i=1}^{N} x_i^2$, is (the square of) the $N$-dimensional coordinate radius in hyper-spherical coordinates.
Assuming that this new scaling in $r_N$ is the only modification of the waveform when it passes from 4 to $D$ dimensions, far away from the source, but still at non-cosmological distances (the so-called wave zone), the binary waveform \eqref{eq:wf_s} will scale as\footnote{In order to recover the right dimensional units in Eq.~\eqref{eq:wf_sD} one should introduce an integration constant as obtained from the solution \eqref{eq:003}. However since this constant can be defined as the scale of transition between a 4 and $D$ dimensional spacetime (see below), we will leave it undefined for the moment being.}
\begin{equation}
	h_\times (t_s) \propto \frac{4}{r_N^{(D-2)/2}} \left( G M_c \right)^{5/3} \left( \pi f_s \right)^{2/3} \cos\theta \sin\Phi(t_s) \,.
	\label{eq:wf_sD}
\end{equation}
As mentioned before, however, higher dimensions are effective only at cosmological distances and thus we need to take as well into account cosmological effects in the waveform~\eqref{eq:wf_sD}.
To do this we repeat the steps that usually lead to \eqref{eq:wf_o}; see e.g.~\cite{Maggiore:2007ulw}.
We start by generalizing the FRW metric to $D$-dimensions
\begin{equation}
	ds^2 = -dt^2 + a(t)^2 \sum_{i = 1}^{N} dx_i^2
	     = -dt^2 + a(t)^2 \left( dr_N^2 + r_N^2 d\Omega_{N-1}^2 \right) \,,
	\label{eq:FRW_line_element_D}
\end{equation}
where $d\Omega_{N-1}^2$ (see Eq.~\eqref{eq:005}) is the angular line element in $N-1$ dimensions.
A similar calculation to the one presented in Appendix~\ref{app:GW_propag_D}, shows that integrating Eq.~\eqref{eq:001} with the metric \eqref{eq:FRW_line_element_D} gives\footnote{Note that for a higher-dimensional FRW universe described by the metric \eqref{eq:FRW_line_element_D}, the definition of redshift does not depend on the number of dimensions (see Appendix~\ref{app:def_z}).}
\begin{equation}
	\mathcal{A} \propto \left(a(t)r_N\right)^{-\frac{D-2}{2}} \,,
\end{equation}
as one would expect.
We then have that the waveform \eqref{eq:wf_sD} at the observer becomes
\begin{equation}
	h_\times (t_o) 
		\propto \frac{4}{(1+z)(a_0 r_N)^{(D-2)/2}} \left( G \mathcal{M}_{cz} \right)^{5/3} \left( \pi f_o \right)^{2/3} \cos\theta \sin\Phi(t_o) \,.
\end{equation}
We can rewrite this waveform as
\begin{equation}
    h_\times (t_o) = \frac{4}{d_L^{\rm GW}} \left( G \mathcal{M}_{cz} \right)^{5/3} \left( \pi f_o \right)^{2/3} \cos\theta \sin\Phi(t_o) \,,
\end{equation}
where we have defined
\begin{equation}
	d_L^{\rm GW} \propto (1+z)(a_0 r_N)^{(D-2)/2} \,,
	\label{eq:002}
\end{equation}
as the quantity inferred by parameter estimation over the measured GW signal assuming a standard GR template with amplitude inversely proportional to the luminosity distance\footnote{Note that the actual waveform model used to perform parameter estimation over the observed GW signal will differ from the lowest-order one presented in Eq.~\eqref{eq:wf_o}. However the amplitude will always be inversely proportional to the luminosity distance in GR, implying that the arguments exposed here will apply anyway whatever waveform model one considers.}.

In order to compare GW and EM measurements of the luminosity distance, we must find the relation connecting $d_L^{\rm GW}$ to the standard 4-dimensional luminosity distance \eqref{eq:def_dL4} which, assuming that light still propagates in 4-dimensions, is the quantity measured by EM observations, namely $d_L^{\rm EM} = d_L^{(4)}$.
First we consider how the definition of the luminosity distance generalises to $D$ dimensions.
From its definition in terms of observable quantities and from simple geometrical consideration, in a $D$-dimensional FRW universe the luminosity distance reads \cite{1992ApJ...397....1C} (see Appendix~\ref{app:dL}):
\begin{equation}
	d_L^{(D)} = a_0 r_N (1+z)^{2/(D-2)} \,.
	\label{eq:dL_D}
\end{equation}
Note that by direct inspection of Eq.~\eqref{eq:wf_sD} this expression gives the correct scaling for GWs propagating in a $D$ dimensional spacetime, namely
\begin{equation}
	h \propto 1/(d_L^{(D)})^{(D-2)/2} \propto 1 / d_L^{\rm GW} \,,
\end{equation}
as one would expect from Eq.~\eqref{eq:wf_sD}.
To find the relation between $d_L^{\rm GW}$ and $d_L^{\rm EM}$, we need thus to find the relation between $d_L^{(D)}$ and $d_L^{(4)}$.
This can be found by noticing that the coordinate distances travelled by a GW and EM signal emitted at the same time by the same source are indeed the same, since we are assuming they arrive at the same time at the observer (we assume that GWs travel at the speed of light at all frequencies).
In fact both signals travel from the source to the observer along null radial geodesics of their respective FRW spacetimes, implying that $ds_D^2$, $ds_4^2$, $d\Omega_{N-1}^2$, $d\Omega_{2}^2$ all vanish.
From their FRW line elements, integrating from the source to the observer, we thus get
\begin{gather}
\int \frac{dt}{a(t)} = \int dr_N = r_N^{\rm src} \,,\\
\int \frac{dt}{a(t)} = \int dr_3 = r_3^{\rm src} \,,
\end{gather}
which immediately gives $r_N = r_3$ for the GW source.
From the equations above we thus find
\begin{equation}
	d_L^{(D)} = a_0 r_3 (1+z)^{2/(D-2)} = d_L^{(4)} (1+z)^{(4-D)/(D-2)} \,,
\end{equation}
which correctly reduces to an equivalence for $D=4$.
Putting everything together we thus get
\begin{align}
	d_L^{\rm GW} &\propto (1+z)(a_0 r_N)^{(D-2)/2} \,,\nonumber\\
				 &= (1+z)(a_0 r_3)^{(D-2)/2} \,,\nonumber\\
				 &= (1+z)\left(\frac{d_L^{(4)}}{(1+z)}\right)^{(D-2)/2} \,,\nonumber\\
				 &= (1+z)^{(4-D)/2}\left(d_L^{\rm EM}\right)^{(D-2)/2} \,,
	\label{eq:gwemD}
\end{align}
providing the relation we were seeking.
This can be rewritten as
\begin{equation}
	d_L^{\rm GW} \propto d_L^{\rm EM} \left(\frac{d_L^{\rm EM}}{(1+z)}\right)^{(D-4)/2} \,,
	\label{eq:GW-EM_no_screening}
\end{equation}
which for $D=4$ correctly recovers $d_L^{\rm GW} = d_L^{\rm EM}$, assuming the constant of proportionality goes to one as $D\rightarrow 4$.
In order to fix this constant however it is simpler to directly consider
the transition from 4-dimensions at small scales to $D$-dimensions at large scales.

As we mentioned above, in order to satisfy all tests of GR at small scales \cite{Liu:2000zq,Overduin:2000gr}, we must require that below a certain scale the spacetime becomes 4-dimensional.
In practice this means that, if we want a relation between $d_L^{\rm GW}$ and $d_L^{\rm EM}$ valid at all scales, we must introduce a scale below which this relation becomes an identity.
We can thus follow the standard phenomenological approach~\cite{Deffayet:2007kf,Pardo:2018ipy,Abbott:2018lct}, in which a distance scale $R_c$ is directly introduced in the relation \eqref{eq:GW-EM_no_screening} to separate the small-scale 4$D$ regime to the higher-dimensional large-scale regime, as follows
\begin{equation}
	\boxed{d_L^{\rm GW} = d_L^{\rm EM} \left[1 + \left( \frac{d_L^{\rm EM}}{R_c (1+z)} \right)^n \right]^{(D-4)/(2n)}} \,.
	\label{eq:dL_z_relation}
\end{equation}
This correctly reduces to an identity for $d_L^{\rm EM} \ll R_c (1+z)$ and to Eq.~\eqref{eq:GW-EM_no_screening} for $d_L^{\rm EM} \gg R_c (1+z)$, where now $R_c$ defines the constant of proportionality, and the constant $n$ determines the steepness of the transition from the small-scale to large-scale behaviour.
Of course we could have chosen a different function to describe the transition from 4-dimensions to $D$-dimensions, but in our phenomenological approach we do not worry too much about the actual form of this transition as long as the two interesting regimes are recovered at small and large scales.
Eq.~\eqref{eq:dL_z_relation} is the expression we need to use when comparing luminosity distance measurements obtained from multi-messenger data!
Note that the factor $(1+z)$ within the square brackets, appears for the first time here and has always been neglected before in the literature; see e.g.~\cite{Pardo:2018ipy,Abbott:2018lct,Corman:2020pyr}.
This factor has the physical effect of redshifting the distance scale at which the transition to higher dimensions takes place.
As we will see in what follows, it strongly impacts constraints on higher dimensional cosmologies obtained from GW+EM multi-messenger events at high-redshift.

\section{Bayesian inference method}
\label{sec:bayesian}
In this paper, we use a Bayesian framework to infer the dimension $D$ and the distance scale $R_c$ from the multi-messenger observation of an astrophysical merger event. Given a dataset $x = n + h(\blmb)$, containing noise $n$ and some signal $h(\blmb)$ modelled using a parameter set $\blmb$, one can obtain a \emph{posterior} probability distribution on $\blmb$, $p(\blmb | x)$, using the Bayes theorem:

\begin{equation}\label{eq:bayes}
    p(\blmb | x) = \frac{\theta(x) \mathcal{L}(x | \blmb)}{E(x)}
\end{equation}
where $\theta(x)$ is the \emph{prior} probability distribution, $\mathcal{L}(x | \blmb)$ is the \emph{likelihood} function, and $E(x)$ is a normalisation constant also called the \emph{marginal likelihood} or the \emph{evidence}. If one is interested in a subset of the parameter set $\blmb$, say $\bxi$, one can marginalise over the complementary set of \emph{nuisance} parameters, $\bxic$, as:

\begin{equation}
    p(\bxi | x) = \int p(\blmb | x) d\bxic
\end{equation}
Given measurements of the GW luminosity distance, $\dgw$, EM luminosity distance, $\dem$ and the redshift, $z$ from statistically independent datasets $\xgw$, $\xem$ and $\xz$ respectively, one can use the joint posterior probability distribution, $p(D, R_c, \dgw, \dem, z|\xgw, \xem, \xz)$ to provide constraints on $D$ and $R_c$, by marginalising over the nuisance parameters. For this, we use two complementary expressions for the posterior. In the first method, we use the samples-based approach introduced in~\cite{Abbott:2018lct} but, for the first time, including an independent measurement of the redshift, $z$, and the new redshift factor as outlined in Eq.~\eqref{eq:dL_z_relation}. The 1D marginalised posterior on $D$ is:
\begin{equation}
\begin{split}
    p(D|\xgw, \xem, \xz) = \int \int \int d\dgw~d\dem~dz~p(\dgw|\xgw)~p(\dem|\xem)~p(z|\xz) \\
    \delta[D - D(\dgw, \dem, z, \rc, n)] \label{eq:samples_based_post_3D}
\end{split}
\end{equation}
The posterior of $\rc$ is identical to the above equation, with $D$ and $\rc$ exchanged. A more likely observational scenario, as outlined in~\cite{Abbott:2018lct,Graham:2020gwr}, is one where we can only measure two, rather than three, of the observables $\{\dgw, \dem, z\}$ independently and infer the third, assuming a model for the expansion of the Universe. This is especially true for LISA sources, as we discuss in detail in Sec.~\ref{sec:LISA}, where we are expected to have measurements of only $\dgw$ and $z$, and not of $\dem$. For such a two-dimensional reduced problem, we can infer $\dem=\dem(z)$ assuming a standard LCDM model for the expansion of the universe. Consequently, the posterior on $D$ can be written as:
\begin{equation}
    p(D|\xgw, \xz) = \int \int d\dgw~dz~p(\dgw|\xgw)~p(z|\xz)~\delta[D - D(\dgw, \dem(z), z, \rc, n)] \label{eq:samples_based_post_2D}
\end{equation}
Just like Eq.~\eqref{eq:samples_based_post_3D}, the posterior on $\rc$ can be obtained by exchanging the positions of $D$ and $\rc$. We use these expressions to provide constraints on the LIGO-Virgo events, GW170817 and GW190521 in the next section (Sec.~\ref{sec:constraints}).

For LISA sources we use a nested sampling algorithm \texttt{Nestle} \cite{2021ascl.soft03022B} instead. This allows us to not only compute the joint posterior on $D$ and $R_c$ but also the evidence which we need for model comparison. Starting from Bayes theorem \eqref{eq:bayes} and assuming we observe a set of $n$ MBHB merger events measured at redshifts $\mathbf{\xz}=\{ {\xz}_1,\cdots {\xz}_n\}$ and GW distances $\mathbf{\xgw}=\{ {\xgw}_1,\cdots {\xgw}_n\}$ 
then provided the measurement uncertainties on the sirens are all independent, the likelihood of the observed data can be written as
\begin{equation}\label{eq:lkh}
p(\mathbf{\xz},\mathbf{\xgw} | D,R_c)= \prod_{i=1}^n \,
p({\xz}_i , {\xgw}_i | D,R_c) 
\end{equation}
Now applying Bayes theorem consecutively the likelihood for a single event can be written as
\begin{equation}
p({\xz}_i , {\xgw}_i | D,R_c) = \int p({\xgw}_i , {\xz}_i , z_i | D,R_c) dz_i  = 
\int p({\xgw}_i | z_i , D,R_c) p ( {\xz}_i | z_i) p(z_i) dz_i ,
\end{equation}
If we further assume that each measured redshift is subject to an independent, normally distributed uncertainty $u_i \sim {\cal N}(0,\Delta_i)$ and assuming that the width of the Gaussian $p ( {\xz}_i | z_i)$ that describes the redshift uncertainties is small compared with the scale over which the distribution of true redshifts is varying, we can approximate $p(z_i)$ in equation above as a constant, so that the marginalisation integral simplifies to
\begin{equation}
p({\xz}_i , {\xgw}_i | D,R_c) = {\cal C} \int p({\xgw}_i | z_i , D,R_c) p ( {\xz}_i | z_i) dz_i ,
\end{equation}
where to be conservative we evaluate the integral over $z_i$ over the range 
$({\xz}_i - 5\Delta_i , {\xz}_i + 5\Delta_i)$ and the normalisation constant is independent of the parameters we are trying to model. This completes the derivation of the likelihood . 
Finally given some prior on the model parameters $p(D,R_c)=p(D)p(R_c)$ which we discuss in Sec.~\ref{sec:LISA} we have all the ingredients to compute the joint or marginalised posteriors on $D$ and $R_c$.
We use these expressions to forecast the ability of LISA to provide constraints on future MBHB events.


\section{Constraints from current ground-based interferometers}
\label{sec:constraints}

The LIGO--Virgo detectors~\cite{TheLIGOScientific:2014jea,TheVirgo:2014hva}, over their three observing runs, O1/O2/O3, have observed two GW events with (potential) EM counterparts: GW170817~\cite{TheLIGOScientific:2017qsa} and GW190521~\cite{Abbott:2020tfl}. GW170817 was followed up by more than 70 terrestrial and space-based EM observatories~\cite{GBM:2017lvd}, which traced its source to the host galaxy, NGC 4993. A claim that AGN J124942.3+344929 (at $z = 0.438$) was a possible EM counterpart of GW190521 was made in~\cite{Graham:2020gwr}\footnote{The claim was made based on a preliminary trigger from the LIGO-Virgo collaborations, S190521g. Following the public announcement of the results of the LIGO-Virgo data analysis~\cite{Abbott:2020mjq}, there has been follow-up work~\cite{Ashton:2020kyr} which has claimed insufficient evidence to confirm AGN J124942.3+344929 as an actual EM counterpart to GW190521. The inconsistency of their final EM and GW luminosity distance measurements shows up in our analysis through a posterior  on $D$ which peaks away from $4$, the nominal GR value. This is likely an artefact of the inconsistent measurements, rather than any significant deviation from GR.}. Joint observations in the GW and EM sectors have allowed GW standard siren~\cite{StandardSirens} measurements of the Hubble constant for both events~\cite{Abbott:2017xzu,Mukherjee:2020kki,Chen:2020gek,Gayathri:2021isv}. GW170817 and GW190521 observations have already been used to constrain extra-dimensions~\cite{Abbott:2018lct,Pardo:2018ipy,Mastrogiovanni:2020mvm}, though without the redshift factor introduced in Eq.~\eqref{eq:dL_z_relation}. In this paper, for the first time, we report constraints on $D$ and $\rc$ for the GW events, GW170817 and GW190521 by including a statistically independent measurement of the redshift $z$, and by considering the new relation given by Eq.~\eqref{eq:dL_z_relation}.

A GW measurement of the luminosity distance for GW170817, $\dgw = 40 ^{+8}_{-14}$ Mpc, was reported in~\cite{TheLIGOScientific:2017qsa}.
A simultaneous measurement of surface brightness fluctuations of the host galaxy, NGC 4993, yielded an EM luminosity distance, $\dem \sim \mathcal{N}(40.7, 2.4 )$ Mpc~\cite{Cantiello:2018ffy}.
These $\dgw$ and $\dem$ measurements are identical to the ones considered in~\cite{Abbott:2018lct}. For our independent redshift measurement, we use the MUSE/VLT measurement reported in~\cite{Hjorth:2017yza}: $z_{\text{cosmic}} = 0.00980 \pm 0.00079$. Following \cite{Hjorth:2017yza}, and in keeping with the measurement in $\dem$, we assume $z$ to be distributed as a Gaussian with mean $0.00980$ and standard deviation $0.00079$. Note this is the predicted redshift due to the Hubble expansion at the distance of NGC4993, the host galaxy of GW170817, after correcting for the (significant) peculiar velocity of NGC4993 -- which can be estimated from a reconstruction of the large-scale peculiar velocity field in its vicinity, as e.g. derived from analysis of all-sky galaxy redshift surveys combined with redshift-independent distance indicator information \cite{Mukherjee:2019qmm,Nicolaou:2019cip}. The standard deviation assigned to $z_{\rm cosmic}$ is based on a conservative estimate of the error associated with deriving this reconstructed peculiar velocity field; see \cite{Abbott:2017xzu} and \cite{Hjorth:2017yza} for further details.
Finally we ignore weak lensing effects on all these measurements; although weak lensing will provide additional systematic uncertainties, these are negligible at low redshift, such as that associated with GW170817.

\begin{figure}[ht]
    \centering
    \includegraphics[width=0.5\textwidth]{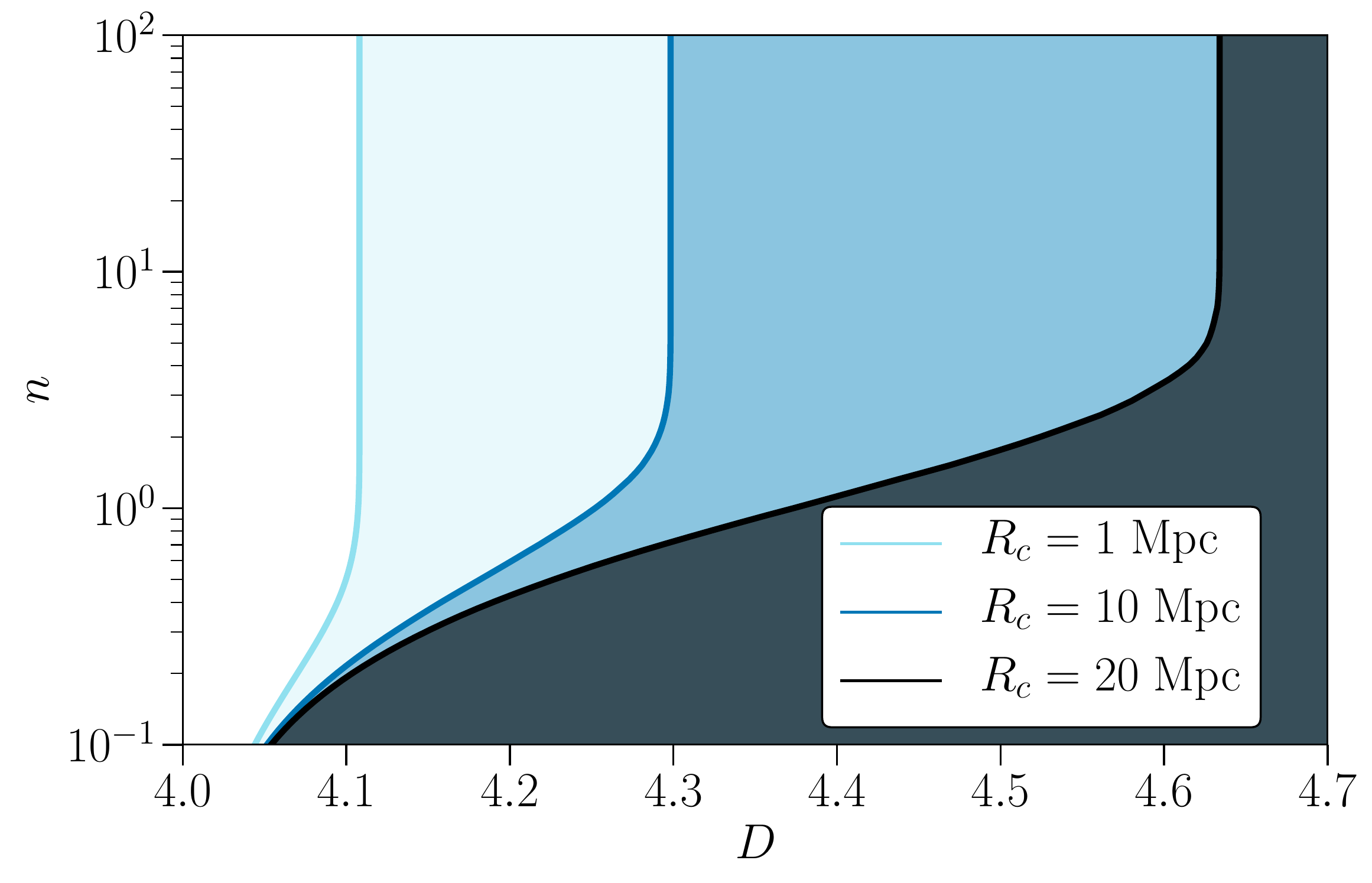}\includegraphics[width=0.5\textwidth]{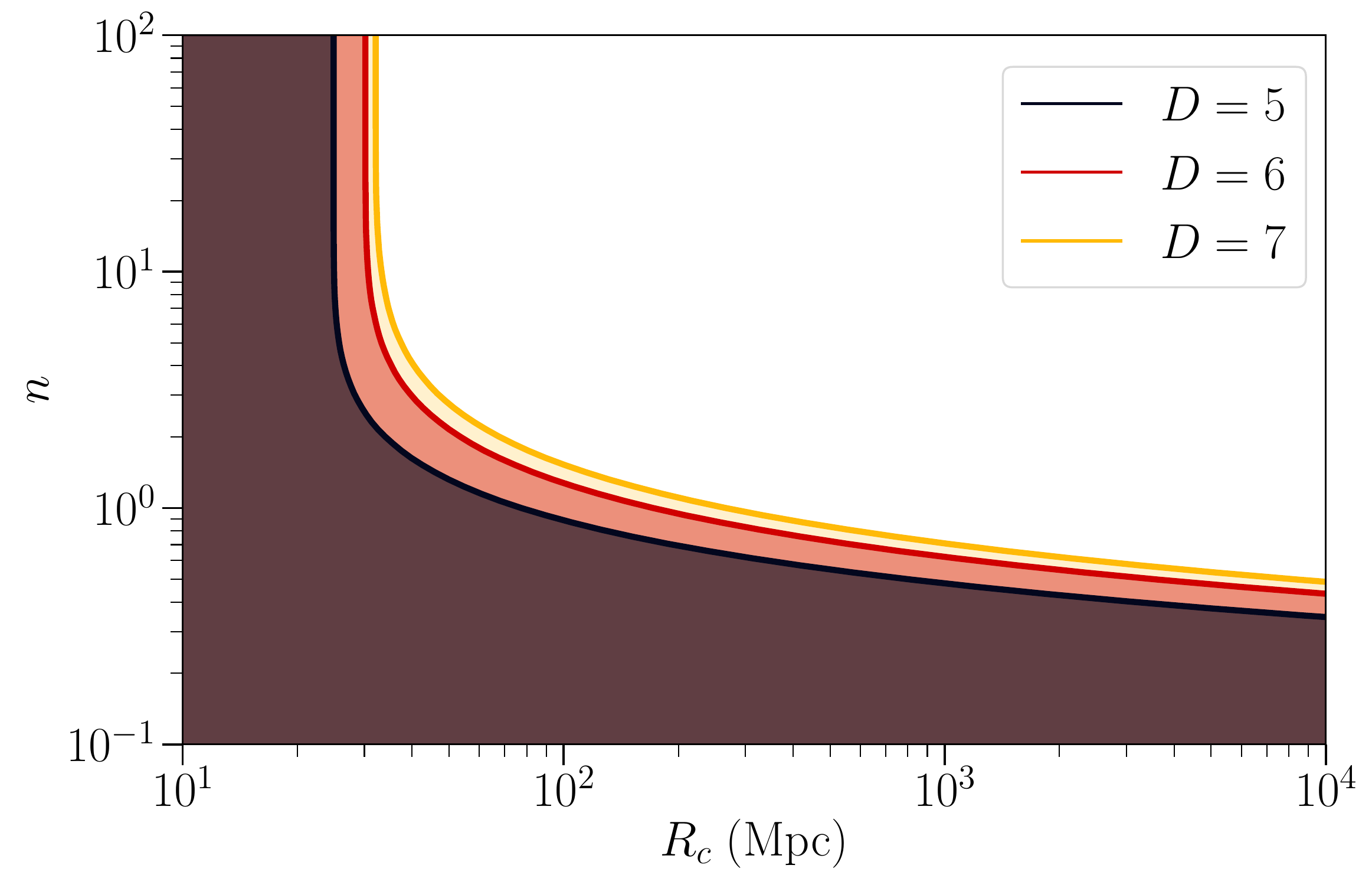}
    \caption{\emph{Left (right) panel}: 95\% upper bounds (lower limits) on the number of spacetime dimensions $D$ (distance scale, $\rc$), assuming fixed transition steepness $n$ and $\rc$ ($D$) for GW170817 with three independent measurements of $\dgw, \dem$ and $z$. Shading indicates the regions of parameter space excluded by the data.}
    \label{fig:gw170817_dlgw_dlem_z}
\end{figure}

The posterior probability distributions on $D$ and $\rc$ assuming three statistically independent measurements of $\dgw, \dem, z$ are inferred using Eq.~\eqref{eq:samples_based_post_3D}. In order to plot the 90\% credible upper (lower) bounds on the posteriors on $D$ ($\rc$) for a given $n$, we choose the same fixed values of $\rc$ ($D$) as in ~\cite{TheLIGOScientific:2017qsa}. The results are plotted in Fig.~\ref{fig:gw170817_dlgw_dlem_z}. We assume that EM radiation always propagates in 4 dimensions, and hence $\dgw > \dem$ or $D>4$. This effectively imposes a prior $D \geq 4$ when estimating $p(D|\xgw,\xem,\xz)$ in the left plot of Fig.~\ref{fig:gw170817_dlgw_dlem_z}. This is also the reason we choose representative values for $D > 4$ i.e., $=5,6,7$ in inferring the lower bounds on $\rc$ in the right plot of Fig.~\ref{fig:gw170817_dlgw_dlem_z}. The fixed values of $\rc$ chosen to infer $p(D|\xgw,\xem,\xz)$ are comparable to the source distance ($\sim 40$  Mpc).

The addition of a third independent variable, $z$, which would consequently need to be marginalised over increases the statistical uncertainty in the measurements of $D$ and $\rc$ compared to those reported in ~\cite{TheLIGOScientific:2017qsa} (specifically, look at Figs. 3 and 4 of ~\cite{TheLIGOScientific:2017qsa}). At the same time, the two results are extremely close, since GW170817 was observed in the very local universe ($z \sim 0.01$). At such distances, the redshift measurement is precise and does not add substantial uncertainty over the measurement uncertainties of $\dgw$ and $\dem$. Similarly, when we consider independent measurements in just $\{\dgw, z\}$, and assume a LCDM model of the universe to infer $\dem$ as a function of the redshift, $z$ \footnote{We use results from the Planck 2018 data release~\cite{Planck:2018vyg}: $\Omega_m = 0.3087; \Omega_\Lambda = 0.6913; h = 0.6764; w_0 = -1; w_a = 0$.\label{fn:cosmovalue}}, the results only marginally improve because of the reduced dimensions of the problem, compared to~\cite{TheLIGOScientific:2017qsa} or the case with three independent variables shown in Fig.~\ref{fig:gw170817_dlgw_dlem_z}. This is because of the lack of measurement uncertainties associated with a third observable. Hence, to summarise, because of the proximity of the GW170817 source, the presence or absence of a redshift measurement does not significantly affect our bounds on $D$ or $\rc$.

\begin{figure}[ht]
    \centering
    \includegraphics[width=0.5\textwidth]{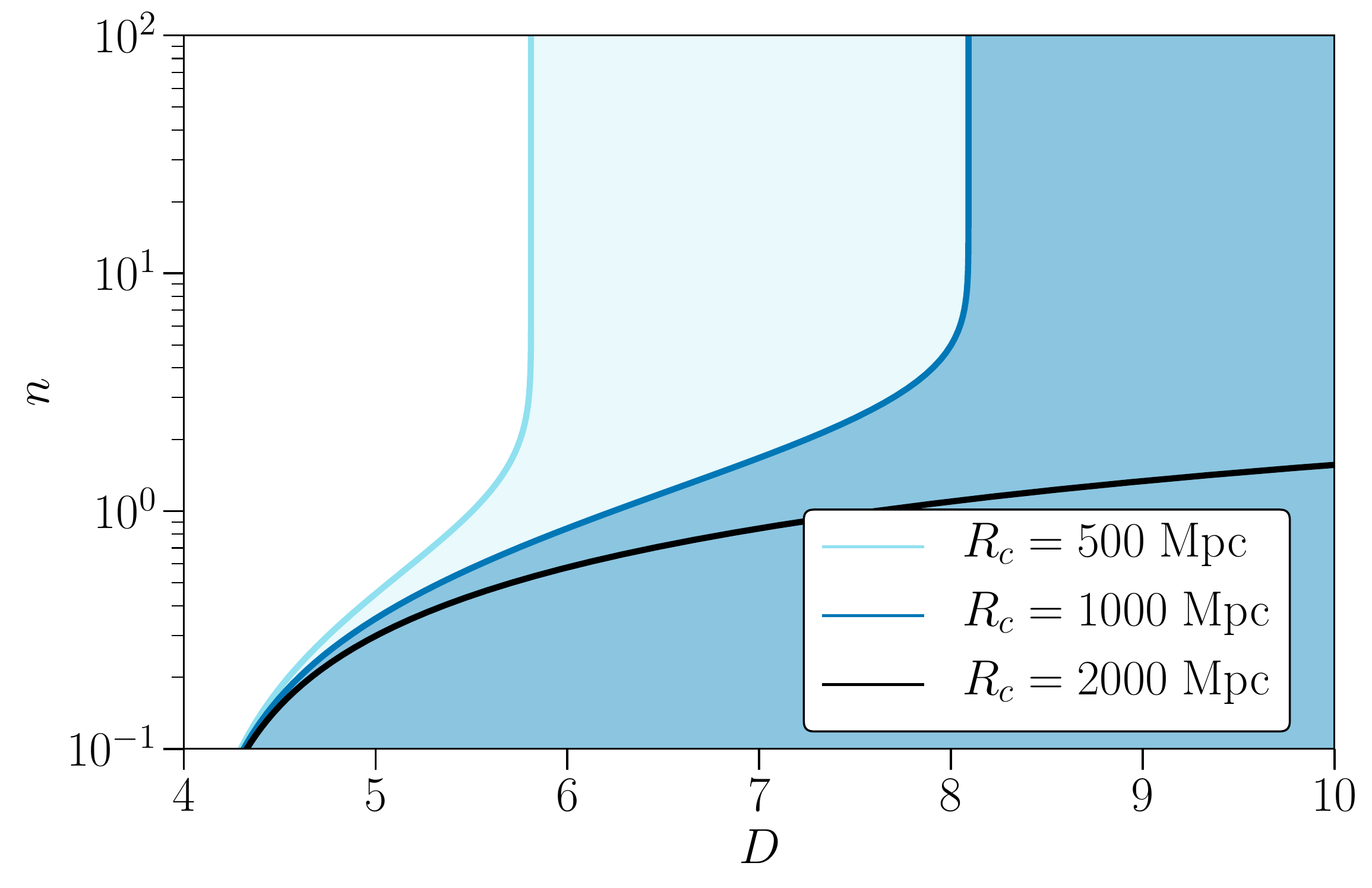}\includegraphics[width=0.5\textwidth]{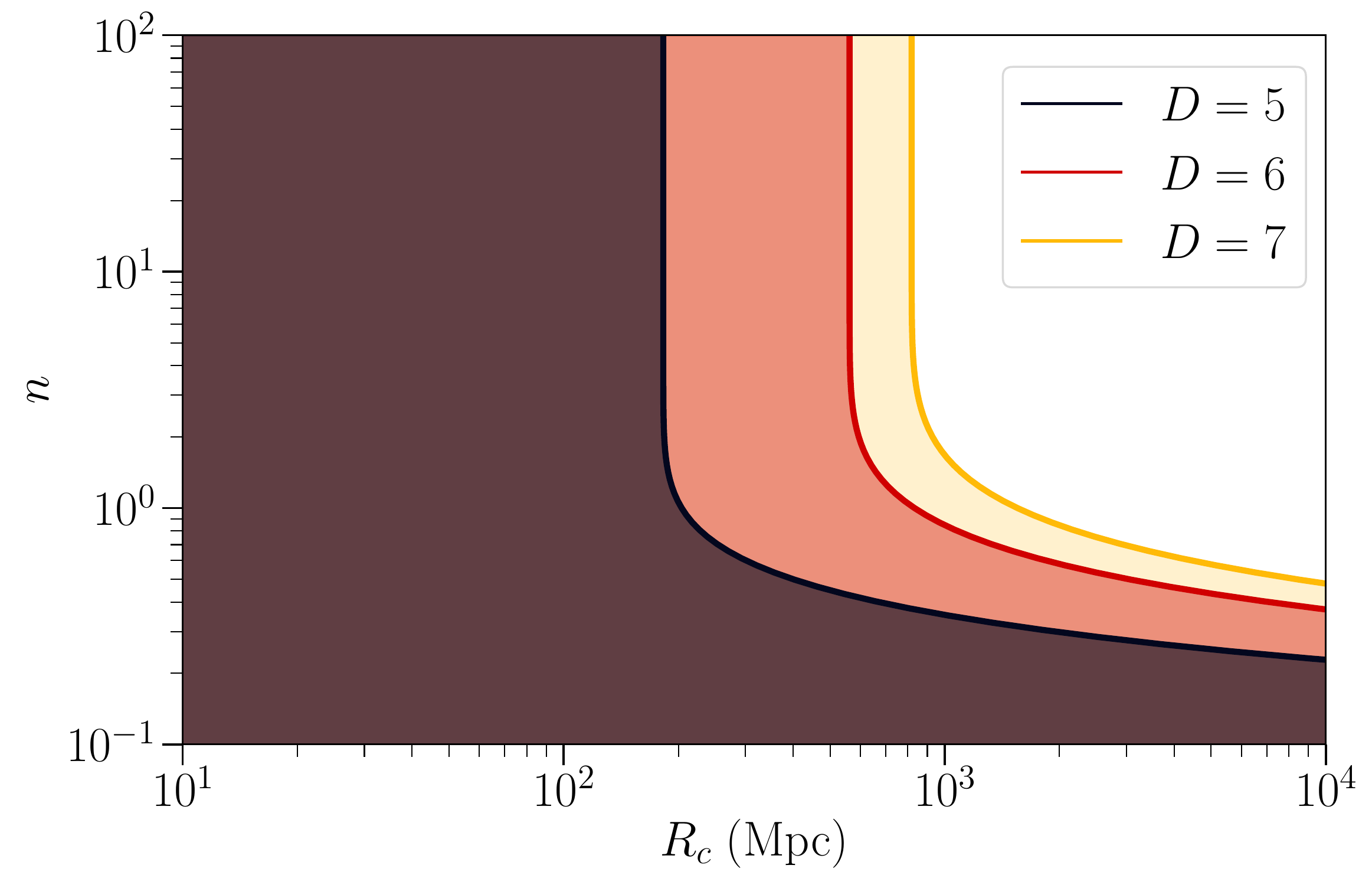}
    \caption{\emph{Left (right) panel}: 95\% upper bounds (lower limits) on the number of spacetime dimensions $D$ (distance scale, $\rc$), assuming fixed transition steepness $n$ and $\rc$ ($D$) for GW190521 with three independent measurements of $\dgw$ and $z$, and assuming a functional relationship between $\dem$ and $z$. Shading indicates the regions of parameter space excluded by the data.}
    \label{fig:gw190521_dlgw_z}
\end{figure}

GW190521 is a slightly different case study. For its GW luminosity distance, we use the measurements with the surrogate numerical relativity binary black hole waveform model, \texttt{NRSurPHM}, reported in~\cite{Abbott:2020tfl}; $\dgw = 5.3^{+2.4}_{-2.6}$ Gpc. This makes GW190521 significantly farther away than GW170817. At such distances, we would expect a larger effect of the redshift measurement on $\{D. \rc\}$. We use the redshift measurement for AGN J124942.3+344929 reported in the SDSS catalog~\cite{SDSS:2017yll}: $z = 0.438 \pm 0.00003$.
We report our bounds in Fig.~\ref{fig:gw190521_dlgw_z}. Just like in the case of  GW170817 above, we restrict ourselves to a prior on $D \geq 4$, and use the same fixed values $5,6,7$ to infer bounds on $\rc$. However, unlike the case of GW170817, since GW190521 was observed at a much larger distance, we consider fixed $\rc$ values comparable to the source distance, i.e.,  500 Mpc, 1 Gpc and 2 Gpc respectively.
We find that, compared to GW170817, GW190521 is able to constrain $R_c$ a lot better at large distances. However, the GW190521 bounds for $D$ are significantly worse. This is unsurprising considering GW190521 was significantly less loud compared to GW170817.

\section{Forecasts for future space-based interferometers: LISA} 
\label{sec:LISA}

In this section we explore how future observations carried out with LISA will help to constrain the number of spacetime dimensions on cosmological scales.
Among all the LISA GW sources that will convey cosmological information (see e.g.~\cite{Kyutoku:2016zxn,DelPozzo:2017kme,Laghi:2021pqk}), the most interesting ones for our scope are mergers of massive black hole binaries (MBHBs) with an identified EM counterpart.
These high-redshift multi-messenger events can in fact efficiently test the expansion of the universe at redshift up to $z\sim 8$ \cite{eLISA,Speri:2020hwc} and probe deviations from the standard cosmological models, namely $\Lambda$CDM, at cosmological epochs still scarcely probed by EM observations \cite{eLISA2,Cai:2017yww,Caprini:2016qxs}.
In what follows we consider simulated catalogs of LISA MBHBs with EM counterparts, and apply the Bayesian approach outlined in Sec.~\ref{sec:bayesian} to derive realistic forecasts on the higher-dimensional cosmological model presented in Sec.~\ref{sec:theory}.

In order to produce catalogs of MBHBs detected by LISA, we closely follow the strategy adopted in \cite{Corman:2020pyr}, which in turn is based on \cite{models,eLISA}, with few minor improvements.
Here we outline only the main details of these catalogs, referring the reader to \cite{models,eLISA} for more information.
The cosmological evolution, merger rates and properties of MBHBs, over which the catalogs depend, are based on the semi-analytic galaxy formation models of \cite{simulation} (see also \cite{Dayal:2018gwg} for a more recent analysis).
Following \cite{models,eLISA} we consider three possible populations of MBHB mergers, based on different underlying astrophysical properties, mainly distinguished by the seeding process that sparks the growth of MBHs over the cosmic history:
\begin{itemize}
    \item \textbf{Model popIII}: A ``light-seed'' scenario where the first massive BHs form from the remnants of population III stars (popIII) \cite{popIII1,popIII2,delays}.
    \item \textbf{Model Q3d}: A ``heavy-seed'' scenario where the first massive BHs form from the collapse of protogalactic disks \cite{heavy1, heavy2, heavy3}, which includes a delay between the coalescence of MBHB host galaxies and that of the MBHs themselves \cite{delays}.
    \item \textbf{Model Q3nod}: Same as Q3d, but assuming no delay between the merger of host galaxies and that of the MBHs.
\end{itemize}
The catalogs contain all information on the MBHBs (their intrinsic properties) and of their astrophysical environment at merger.
These inputs are then used to estimate both the detectability with LISA and the EM counterpart emission.

To check whether any event in each catalog is detected by LISA, we employ the approach of \cite{models,eLISA}. This consists of using a Fisher matrix method to calculate the signal to noise ratio and perform parameter estimation over the GW signal, using inspiral-only precessing waveforms, including spin-spin and spin-orbit interactions up to 2 and 3.5 post-Newtonian (PN) orders.
The contribution to the SNR and the parameter estimation coming from the merger and ring-down phases, is then calculated following the phenomenological approach of \cite{eLISA}.
The detectability threshold is then set to SNR$>8$, but taking into account the full inspiral, merger and ring-down signal.
This procedure allows us to find the LISA measurement uncertainties on the luminosity distance and sky-localisation for each MBHB event, which we require for our cosmological analysis.

In order to use these events as standard sirens, we need to collect an associated redshift measurement for each one of them.
This is done by assuming that an EM counterpart is observed and by measuring the redshift of the corresponding host galaxy.
To estimate the emission and observation of EM counterparts, we follow \cite{eLISA}, to which the reader is referred for full details.
First only MBHB events with a sufficiently accurate sky localisation ($\Delta\Omega < 10\, {\rm deg}^2$) are considered, in order to be able to efficiently point and use realistic future EM facilities.
Then, by using the information on the host galaxy and MBHB environment given by the simulated catalogs, the emission of EM radiation in the optical and radio bands at merger and after-merger is estimated from theoretical models of EM production from MBHBs \cite{Palenzuela:2010nf}.
Two strategies are then employed to check the detectability of the emitted EM radiation and to measure its redshift.
The first one is the detection of an optical flare at merger with LSST\footnote{http://www.vso.org}, the second one is the detection of a post-merger radio jet or flare with SKA\footnote{http://www.skatelescope.org}, followed by a redshift measurement with ELT\footnote{https://elt.eso.org}.
Given its nature, which could give rise to a radio transient lasting up to months after merger, and its higher magnitude, the emission in the radio and detection with SKA is by far the most promising strategy to collect an associated redshift measure, as shown in \cite{eLISA}.

The redshift measurement uncertainty is then estimated as follows, depending whether the measurement is made using spectroscopic or photometric methods (which depends on the properties of the host galaxy, mainly its distance and luminosity; see \cite{eLISA}).
For spectroscopic measurements we assume a constant 1$\sigma$ relative error of 0.01.
For photometric measurements instead we assume that relative 1$\sigma$ measurement errors scale as $0.03 (1+z)$. This corresponds to the ``optimistic'' scenario of \cite{eLISA}.

Finally in order to complete our catalogs we need to consider also the systematic effects of weak lensing and peculiar velocities, which degrade the uncertainty on the luminosity distance retrieved by LISA.
In order to estimate the contribution of peculiar velocities we follow again the expression provided in \cite{eLISA}, which however is not particularly relevant for high-redshift events.
More important is the contribution of weak lensing, which is expected to be the dominant source of uncertainty on the luminosity distance at high-redshift.
In order to estimate this contribution, we use the recent results of \cite{Cusin:2020ezb}, which provides fitting formulas for the average weak lensing uncertainty as functions of redshift.
We also assume that de-lensing is possible up to $z=2$ to a maximum of 30\%, exactly as considered in \cite{Speri:2020hwc}, and effectively we use the same de-lensing model adopted there. 
The peculiar velocities and remaining weak lensing uncertainties are then added in quadrature to the LISA measurement error to provide the total uncertainty on the luminosity distance retrieved from MBHBs.
By following the procedure outlined here above, we produce catalogs of MBHB events, assuming a $\Lambda$CDM cosmology, for which both the luminosity distance and the redshift is measured. Equipped with these mock catalogs we use the functional relation describing the effect of gravitational damping on the GW waveform in higher dimensional theories with non-compact extra dimensions, Eq.~\eqref{eq:dL_z_relation}, to generate 
realisations of MBHB sirens for various possible higher-dimensional theories.
To do so we assume a stationary Gaussian noise $\sigma_i$ on the measured gravitational wave luminosity distance given by procedure outlined above. This could in principle results in MBHBs, in particular those at high redshift, to drop below the adopted SNR threshold for observable events. The effect of this sample correction on the parameter estimation was investigated in \cite{Corman:2020pyr} and was found to be small, hence we do not consider this possible selection bias further. It is also important to note that to be fully self-consistent, we should compute the MBHB merger rates and redshift distributions in our particular chosen higher dimensional theory and not in $\Lambda$CDM. However, we do not expect the rates and distributions obtained in that manner to be significantly different, since the dominant effect is instead the details of the galaxy formation and evolution model adopted. Thus, we adopt the merger rates and redshift distribution calculated for the $\Lambda$CDM model. In what follows we consider the following, arbitrary but not ruled by observations, "injection" model $\{D=5, R_c H_d=1, n=1 \}$ where $H_d=H_0^{-1}$ is the current Hubble radius.

We create 22 catalogs of 4-year LISA observations of MBHBs for each one of the three MBH population models (popIII, Q3d, Q3nod).
These catalogs contain the same number of sources, roughly $\sim$14 events for popIII and Q3d and $\sim$28 for Q3nod, as the ones considered in \cite{Corman:2020pyr} to which we refer for more details.
The nested sampling implementation described in Sec.~\ref{sec:bayesian} is then run on the simulated GW and EM data for each of these 22 catalogs and for the cosmological model presented in Sec.~\ref{sec:theory}.
We assume each measured redshift and gravitational distance is subject to an independent, normally distributed uncertainty such that ${\xz}_i \sim {\cal N}(z_i,\Delta_i)$ 
and $ {\xgw}_i \sim {\cal N}({d^L}_{GW,i},\sigma_i)$ respectively.  
In order to represent complete ignorance about the parameters defining the higher-dimensional theory we take uniform uninformative priors in the range $D \in [4,11]$ and $R_c \in [20,\infty)$ where the lower limit on the screening scale is set by distances ruled out by GW170817 in Sec.~\ref{sec:constraints}. The lower limit on $D$ is chosen to be consistent with section~\ref{sec:constraints} and the upper limit chosen to limit the computational cost of the parameter estimation method.
For the  background cosmology we assume the same cosmology as the one used to generate the catalogs; see footnote \ref{fn:cosmovalue}.
We discuss the results of the parameter estimation for each functional relation on the luminosity distance and consider a model comparison to the $\Lambda$CDM model.

\begin{figure}[t]
    \centering
    \includegraphics[width=0.61\textwidth]{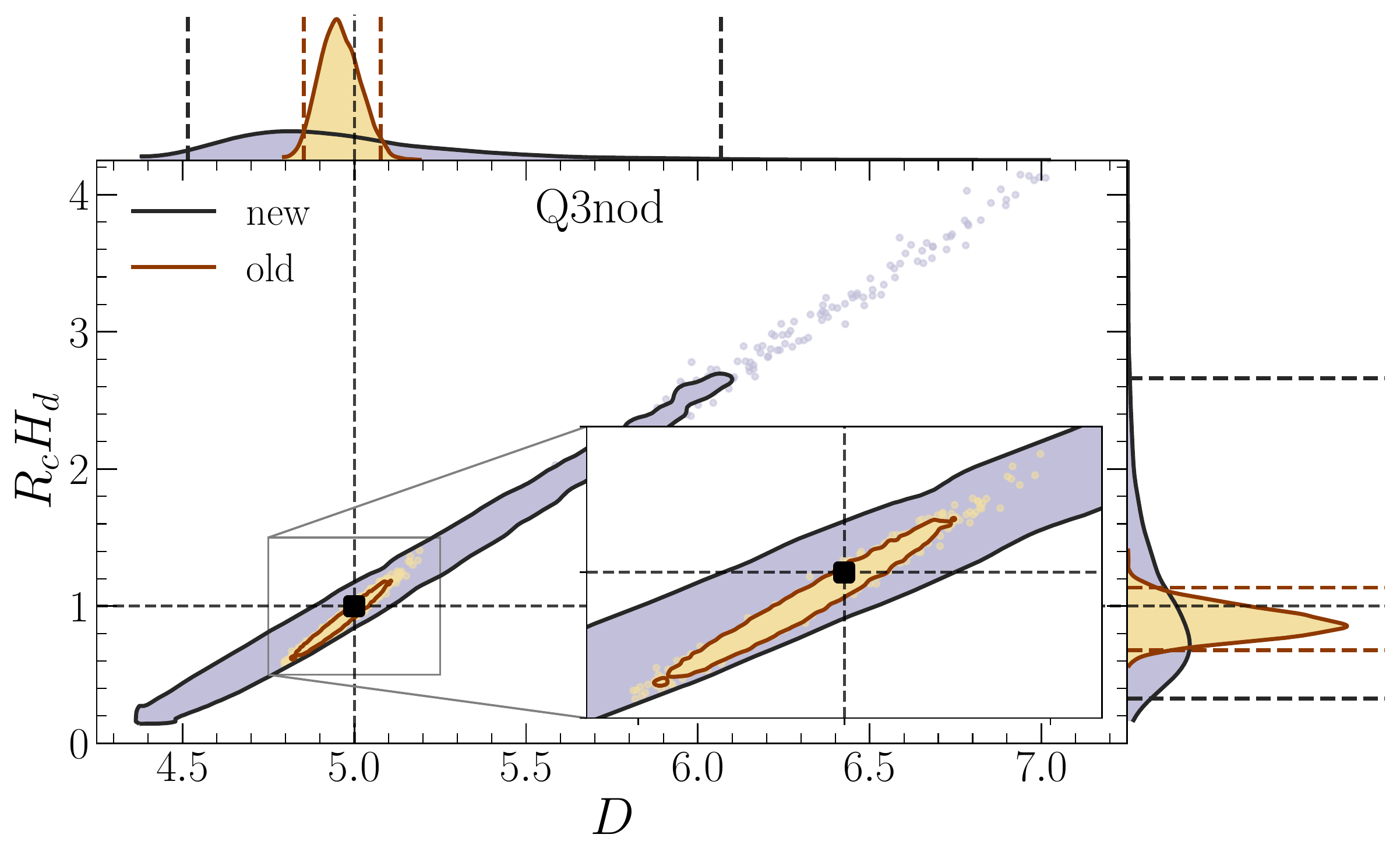}
    \includegraphics[width=0.61\textwidth]{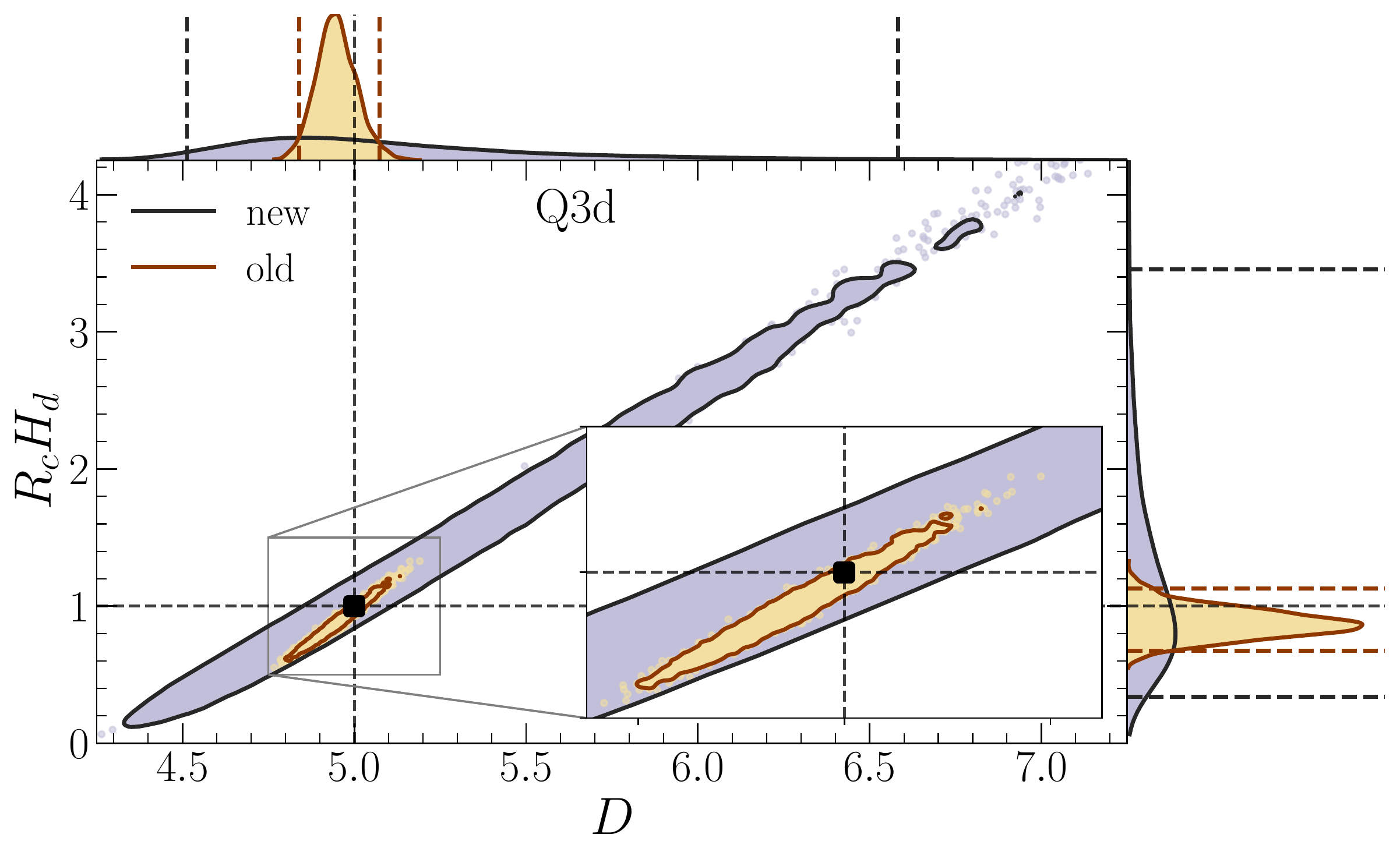}
    \includegraphics[width=0.61\textwidth]{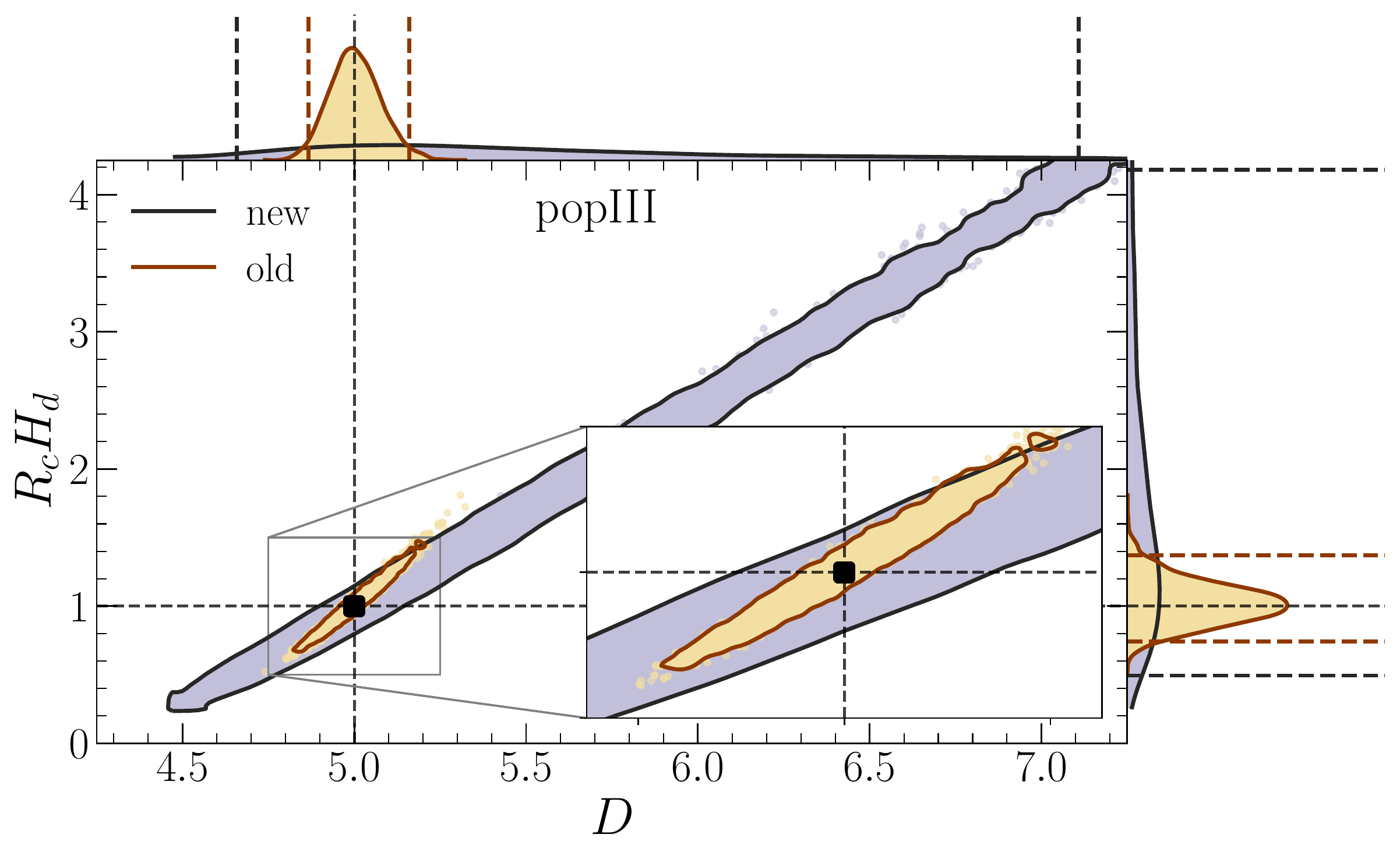}
    \caption{Joint posterior distributions on the number of spacetime dimensions, $D$ and screening scale $R_c$ for the cosmological scenario $(D=5,R_c H_d =1.0)$ and different MBHB formation models using Eq.~\eqref{eq:dL_z_relation} with ("new"/grey) and without ("old"/yellow) the redshift dependence. From top to bottom the formation models considered are: heavy seeds without delays (Q3nod),  heavy seeds with delays (Q3d) and light-seeds (popIII). The (blue/red) dashed lines in the 1D marginalised posteriors indicate the $95 \%$ confidence intervals. The black square and black dashed lines represent the injected values.
    }
    \label{fig:olp_lisa}
\end{figure}

The 2D parameter estimation on $\{D,R_c\}$ described above gives us a joint posterior distribution and 1D posteriors on each parameter for each catalog within a given MBHB formation model. To further quantify the capability of LISA to constrain either $D$ or $R_c$,
we first record the median values of the marginalised posterior of each cosmological parameter for all the catalog
realisations.
We then use the median as a Figure-of-Merit (FoM) for these parameter estimates. For an estimate of the LISA error on the median FoM we first take the 95 $\%$ credible interval (CI) around the median value for each catalog and then adopt the median of these intervals to represent the 95 $\%$ CI of the FoM. When we compare the capability of LISA to place constraints on a given parameter for different MBHB models, we will always use the median FoM together with 95 $\%$ CI, which essentially captures the scatter in the FoM -- and hence provides a realistic estimate of the  expected statistical uncertainty. Note that there is also a significant scatter in the characteristics of the MBHB population between different catalogs. The impact of this scatter on the cosmological constraints that we can place is non-negligible and was studied in \cite{Corman:2020pyr}.
Figure \ref{fig:olp_lisa} shows the joint posterior PDF over our two-dimensional model parameter space $\{D,Rc\}$
for the FoM catalog for the "new" (with the $(1+z)$ factor in in Eq.~\eqref{eq:dL_z_relation}) and "old" (without) functional relation for each MBHB formation model. Table~\ref{lisa_stats} presents FoMs, derived from the marginalised PDFs, for each of the cosmological model parameters, for all MBHB formation scenarios and for each functional relationship. In each entry of Table \ref{lisa_stats}, the top row shows the median FoM and 95 $\%$ CI for light seeds (popIII), the central row for heavy seeds with delays (Q3d) and the bottom row for heavy seeds without delays (Q3nod). We find, in agreement with our previous analysis \cite{Corman:2020pyr}, that Q3nod systematically gives better results than the other two scenarios, which are roughly comparable to each other, due to their lower number of detectable standard sirens. This shows that the extent to which LISA can be used to perform meaningful constraints on theories of modified gravity defined by the new scaling Eq. ~\eqref{eq:dL_z_relation} will still strongly depend on the actual redshift distribution of MBHB merger events and the corresponding efficiency in identifying an EM counterpart. But we also find that the new functional relation Eq.~ \eqref{eq:dL_z_relation} strongly affects our ability to constraint the parameters, details of which we discuss next.

\renewcommand{\arraystretch}{1.9}
\begin{table}[h]
\begin{center}
\caption{\label{lisa_stats} (Left and middle) Median figures of merit and $95\%$ credible intervals summarising the marginalised posterior PDF of $D$, the number of dimensions and $R_c$, the crossover length scale assuming an underlying cosmology with true parameters $\theta=\{D=5,R_c H_d=1,n=1, H_0=67.4\}$, for the different MBHB formation models considered. In each row of the Table separated by dashed lines, the top sub-row shows the FoMs and CI for light-seeds (popIII), the central sub-row for heavy seeds with delays (Q3d) and the bottom sub-row for heavy seeds without delays (Q3nod).
Results denoted by "new" are derived using the modified distance relation taking into account the redshift factor $(1+z)$ in Eq.~\eqref{eq:dL_z_relation}, while results denoted by "old" assume the same relation, without the $(1+z)$ factor, as considered in \cite{Abbott:2018lct,Pardo:2018ipy,Corman:2020pyr}. The right column gives the log Bayes factor for a particular NGR model, compared with the $\Lambda$CDM model. A positive log Bayes factor implies evidence in favour of the NGR model.
}
\begin{tabular}{|c|c|c||c|c||c|c|}
\hline
\multirow{2}{*}{Model} & \multicolumn{2}{c||}{$D$} & 
    \multicolumn{2}{c||}{$R_c H_d$}  & 
    \multicolumn{2}{c|}{$\ln B$} \\
\cline{2-7}
 & old & new & old & new  & old & new \\
\hline
 {popIII}& $5.01^{+0.18}_{-0.14}$  & $5.83^{+1.17}_{-1.04}$&$1.03^{+0.38}_{-0.28}$ & $2.17^{+1.84}_{-1.54}$ &$1074^{+627}_{-672}$ & $141^{+126}_{-101}$  \\
 \cdashline{1-7}
{HQ3} & $5.02^{+0.14}_{-0.13}$  & $5.47^{+1.34}_{-0.75}$&$1.05^{+0.30}_{-0.25}$ &$1.69^{+2.07}_{-1.11}$&$1568^{+722}_{-868}$ & $171^{+125}_{-81}$  \\
\cdashline{1-7}
{HND}& $4.99^{+0.12}_{-0.10}$  & $5.26^{+1.23}_{-0.49}$&$1.01^{+0.25}_{-0.19}$ &$1.35^{+1.87}_{-0.70}$ &$2748^{+629}_{-1217}$ & $321^{+152}_{-145}$  \\
\hline
\end{tabular}
\end{center} 
\end{table}
We first consider the effect of the new functional relation on LISA's ability to place limits on the number of spacetime dimensions.
Figure \ref{fig:olp_lisa} and Table \ref{lisa_stats} both show that, for all formation models, there is a systematic shift in the median value of the number of spacetime dimensions away from its true value (indicated by the black square in the figure)  but also an increase in the statistical spread.
Despite this, the Bayes factors shown in Table \ref{lisa_stats} indicate that we will still be able to distinguish a $\Lambda$CDM universe from a higher dimensional cosmology, at least according to Jeffrey's scale.
Since LISA will be observing events up to high redshifts, unlike LIGO, it is not surprising that the extra redshift dependence in the  GW luminosity distance relation strongly affects the constraints. Again we note that the formation model with the highest number of sources, namely Q3nod, gives the best results.\\
Considering next the constraints of the screening scale, we find a similar trend -- although the effect is more pronounced. Looking at the joint posterior distribution we find that, for the cosmological scenario considered, the number of dimensions and screening scale are strongly correlated. This is not because of the new functional relation for the luminosity distance and was also the case when the redshift factor is ignored. However, it was found in \cite{Corman:2020pyr} that, as the transition to a higher dimensional cosmology becomes steeper and/or the screening scale becomes closer, the parameters become uncorrelated.
For completeness we also considered a scenario where the screening scale is four times the Hubble radius and hence the deviation from GR at small redshift is less pronounced; in this case we find, in agreement with our previous study, that the constraints on the screening scale are less accurate. Similarly if one were to consider a cosmological scenario where the number of spacetime dimensions is greater than five, the constraints would improve due to the more significant deviations from 4 dimensions. We refer the reader to \cite{Corman:2020pyr} for more details on these cases.

\section{Discussion and conclusion}
\label{sec:conclusion}

In this paper we reconsidered the possibility of constraining extra spatial-dimensions at cosmological distances with GW multi-messenger events.
We revised the theory behind the derivation of the relation between $\dgw$ and $\dem$, spelling out all underlying assumptions and explicitly presenting all details of the calculation.
By doing this, we showed that an additional redshift factor, which was neglected in previous analyses, must actually be included in the GW luminosity distance relation given by Eq.~\eqref{eq:dL_z_relation}.
We then revised the constraints derived from current and future GW observations, respectively with LIGO-Virgo and LISA, taking into account the new luminosity distance relation.
To derive constraints on higher dimensions from current GW observations carried out with the LIGO-Virgo detectors, we considered the data analysis strategy of \cite{Abbott:2018lct,Pardo:2018ipy}, and expanded it to include the possibility of an independent measurement of $\dem$.
To produce forecasts for future GW observations taken from space with LISA, we instead considered and expanded the approach of \cite{Corman:2020pyr}. Our two approaches were shown to provide consistent results.

Our results were presented in Sec.~\ref{sec:constraints} and Sec.~\ref{sec:LISA}.
As expected, constraints obtained with low redshift events such as GW170817 do not change appreciably from the ``old'' to the ``new'' luminosity distance relation, i.e.~respectively by neglecting or considering the $(1+z)$ factor in Eq.~\eqref{eq:dL_z_relation}.
The correction due to this factor at low redshift are in fact small, and do not affect the final results as long as sub-percent accuracies are not reached.
For this reason, the constraints presented in \cite{Abbott:2018lct,Pardo:2018ipy} can equally apply to the new relation \eqref{eq:dL_z_relation} without loss of generality.
The constraints that we derived using GW190521 are, however, less impressive due to the large uncertainties associated with the measurements of this event -- although the transition scale $R_c$ can effectively be better constrained due to the larger distance of this GW event.
Note however that the association of an EM counterpart with GW190521 is at best debatable, and constraints derived from this event should not be taken seriously.
As an academic exercise, they show however how higher-redshift multi-messenger GW events can help to  constrain higher-dimensional models, notably by bounding the screening scale to larger distances.

A different conclusion applies, on the other hand, to the constraints derived from our simulated LISA MBHB sources.
At high redshift, effectively at $z\gtrsim 1$, the new $(1+z)$ factor appearing in Eq.~\eqref{eq:dL_z_relation} strongly affects the derived constraints.
As shown in Sec.~\ref{sec:LISA}, expected constraints on the number of spacetime dimensions at cosmological distances effectively worsen by almost an order of magnitude with respect to the results obtained without taking into account the $(1+z)$ factor.
Fortunately, however, this does not hinder the ability of LISA to efficiently constrain these higher-dimensional cosmological models.
In fact, so long as the screening scale $R_c$ is not taken to be larger than the Hubble radius, LISA will be able to distinguish between $\Lambda$CDM and a 5-dimensional cosmological model irrespective of the underlying astrophysical properties of the MBHB population. 
More generally the higher the distance of our detected multi-messenger events, the larger the transition scale $R_c$ that can be constrained. In fact from our analysis it is clear that current LIGO/Virgo results, in particular GW170817, can only constrain $R_c$ up to tens of Mpc, while LISA will allow us to push constraints up to Gpc scales with the detection of MBHB mergers.

We conclude by stressing that the higher-dimensional cosmological spacetime considered in this paper is only a simple phenomenological toy-model that leads to a GW-EM luminosity distance relation expected to well describe better motivated higher-dimensional modified gravity models, used for example to characterise the current acceleration of the universe.
Nevertheless, by excluding the possibility of simple higher-dimensional extensions of the homogeneous and isotropic spacetime that well capture the overall features of our universe at the largest scales, we can help directing the efforts aimed at understanding the fundamental nature of the observed cosmic acceleration, which will then be restricted to consider only models admitting exactly four dimensions.
Of course more complex higher-dimensional models can always be constructed in order to avoid the constraints derived here, but these will only come at the cost of abandoning the theoretical and observational simplicity of a homogeneous and isotropic distribution of matter in the universe.

\acknowledgments
We thank S.~Mastrogiovanni and S.~Mukherjee for useful discussions. 
N.T.~is supported by an ANR Tremplin ERC grant (ANR-20-ERC9-0006-01). CE-R acknowledges the \textit{Royal Astronomical Society} as FRAS 10147 and the financial supported by DGAPA-PAPIIT-UNAM Project IA100220. M. H. is supported by the Science and Technology Facilities Council (Ref. ST/L000946/1). This article is also based upon work from COST action CA18108, supported by COST (European Cooperation in Science and Technology). M.C. is supported by the Perimeter Institute for Theoretical Physics. Calculations were in part performed on the Symmetry cluster at Perimeter Institute.

\appendix

\section{Appendix: details of calculations of Sec.~\ref{sec:theory}}

\subsection{Damping of GWs in higher dimensions}
\label{app:GW_propag_D}

Here we prove that using the Minkowski $D$-dimensional metric \eqref{eq:le_Minkowski_D} in (hyper-)spherical coordinates, namely
\begin{equation}
	ds^2 = -dt^2 + dr_N^2 + r_N^2 d\Omega_{N-1}^2 \,,
	\label{eq:004}
\end{equation}
the integration of Eq.~\eqref{eq:001} yields the scaling \eqref{eq:003} for the GW amplitude.
In the metric \eqref{eq:004} the explicit expression for $d\Omega_{N-1}^2$ is
\begin{equation}
	d\Omega_{N-1}^2 = d\theta_1^2 + \sin^2\theta_1 d\theta_2^2 + \sin^2\theta_1 \sin^2\theta_2 d\theta_2^3 +
		\cdots + \sin^2\theta_1 \cdots \sin^2\theta_{N-2} d\theta_{N-1}^2 \,,
	\label{eq:005}
\end{equation}
where $\theta_i$ are the $N-1$ angular coordinates in the $D=N+1$ dimensional spacetime.
First we note that Eq.~\eqref{eq:001} can be rewritten as
\begin{equation}
	\nabla_\mu \left( \mathcal{A}^2 k^\mu \right) = \frac{1}{\sqrt{-g}} \partial_\mu \left( \sqrt{-g} \mathcal{A}^2 k^\mu \right) \,,
	\label{eq:006}
\end{equation}
where $g$ is the determinant of the $D$-dimensional metric and $\partial_\mu$ is the usual partial derivative.
For the metric \eqref{eq:004} we find
\begin{equation}
	\sqrt{-g} = r_N^{D-2} \sin^{D-3}\theta_1 \sin^{D-2}\theta_2 \cdots \sin^2\theta_{D-3} \sin\theta_{D-2} \,.
\end{equation}
Considering a GW propagating from the origin of the coordinates toward the radial direction, we have
\begin{equation}
	k^\mu = \left( k^0, k^r, 0, \cdots, 0 \right) = \left( \omega, \omega, 0, \cdots, 0 \right) \,,
\end{equation}
where the second equality comes taking into account that $k^0 = \omega$ (the frequency) and $k^\mu k_\mu = 0$.
In what follows we will assume that the frequency does not depend on the spatial spacetime coordinates, although it is still allowed to depend on the coordinate time $t$.
Using these two last relations into Eq.~\eqref{eq:006} one finds
\begin{equation}
	\frac{\partial}{\partial t} \left( \mathcal{A}^2 \omega \right) + \frac{1}{r^{D-2}} \frac{\partial}{\partial r} \left( r^{D-2} \mathcal{A}^2 \omega \right) = 0 \,.
\end{equation}
Assuming $\dot{\mathcal{A}} \gg \mathcal{A}$ and $\dot\omega \ll \omega$, which are both valid within the geometric optics approximation, we obtain
\begin{equation}
	\frac{\partial}{\partial r} \left( r^{D-2} \mathcal{A}^2 \omega \right) = 0 \,,
\end{equation}
which once integrated gives
\begin{equation}
	\mathcal{A} \propto r^{-(D-2)/2} \,,
\end{equation}
where all dependencies upon $t$ and all the other coordinates have been adsorbed in the proportionality relation.
This coincides with the scaling \eqref{eq:003}.


\subsection{Definition of redshift in higher dimensions}
\label{app:def_z}

In this appendix we demonstrate that the redshift in the $D$-dimensional FRW spacetime \eqref{eq:FRW_line_element_D} coincides with the usual one defined in 4-dimensions.
The cosmological redshift can be defined as the ratio between the observed frequency $f_O$ and the emitted frequency $f_E$ from an EM signal
\begin{equation}
	\frac{f_O}{f_E} = \frac{1}{1+z} \,.
	\label{eq:def_z}
\end{equation}
The frequency can be rewritten as the inverse of the time of arrival $\delta t$ of two subsequent wave-crests, namely $f = 1/\delta t$.
Taking an EM ray propagating along the radial direction ($d\Omega_N^2 = 0$ in Eq.~\eqref{eq:FRW_line_element_D}) toward the origin of the coordinates (the observer) we have
\begin{equation}
	ds^2 = 0 \,, \quad\Rightarrow\quad dt = -a(t) dr_N \quad\Rightarrow\quad \int_{t_E}^{t_O} \frac{dt}{a(t)} = \int_{r_E}^{r_O} dr_N \,.
\end{equation}
Differentiating the last equation with respect to time and recalling that comoving coordinate are time independent, one obtains
\begin{equation}
	\frac{\delta t_O}{a(t_O)} - \frac{\delta t_E}{a(t_E)} = 0 \,.
\end{equation}
Recalling that $f = 1/\delta t$, from the definition \eqref{eq:def_z} one obtains the usual relation between the redshift and the scale factor
\begin{equation}
	\frac{1}{1+z} = \frac{a(t_E)}{a(t_O)} \,,
\end{equation}
which is thus valid also in the $D$-dimensional FRW spacetime.
In conclusion the redshift in a higher dimensional FRW universe coincides with the one defined in 4-dimensions.


\subsection{Luminosity distance in higher dimensions}
\label{app:dL}

The luminosity distance can be defined by the relation between the emitted luminosity flux $L_E$ of an astronomical source and the luminosity flux $L_O$ observed by telescopes on the Earth assuming lights propagates spherically in an Euclidean spatial geometry.
In $D = N+1$ dimensions the geometrical relation between $L_E$ and $L_O$ must take into account that the EM flux propagates isotropically in a hyper-sphere embedded in $N+1$-dimensions, implying that the luminosity distance $d_L^{(D)}$ is now defined by the relation
\begin{equation}
	L_O = \frac{L_E}{b_{N-1} (d_L^{(D)})^{N-1}} \,,
	\label{eq:def_D_L}
\end{equation}
where
\begin{equation}
	b_{N-1} = \frac{2 \pi^{N/2}}{\Gamma(N/2)} \,,
\end{equation}
is a geometrical constant factor (the surface of the $N$-sphere of unit radius) and $\Gamma$ is the gamma function.
In the $D$-dimensional FRW spacetime \eqref{eq:004} the ratio between $L_E$ and $L_O$ is given by the surface $S_N$ of a $N$-sphere of radius $r_N$.
This can be computed integrating the line element \eqref{eq:004} at fixed time and radius ($dt = dr_N = 0$)
\begin{align}
	S_N &= \int d\theta_1 \cdots d\theta_{N-1} \sqrt{-g_{N-1}} \,,\\
	    &= \int d\theta_1 \cdots d\theta_{N-1} r_N^{N-1} a(t)^{N-1} \nonumber\\
	       & \hspace{1.7cm} \times \sin^{N-2}\theta_1 \cdots \sin^2\theta_{N-2} \sin\theta_{N-1} \,,\\
	    &= b_{N-1} r_N^{N-1} a(t)^{N-1} \,,
\end{align}
where we used the $N-1$-dimensional angular integration measure $\sqrt{-g_{N-1}}$.
In addition the rate of arrival of individual photons is lower by a factor $1/(1+z)$ due to the cosmological expansion and the energy of each photons is also redshifted by the same factor $1/(1+z)$.
This implies that the luminosity $L_O$ observed from the Earth is lower by the factor $1/(1+z)^2$ because of the cosmological expansion.
Putting all this together we find (note that this relation is evaluated at the observer, so here $a(t) = a_0$; see~e.g.~\cite{Weinberg:2008zzc})
\begin{equation}
	L_O = \frac{L_E}{b_{N-1} (a_0r_N)^{N-1} (1+z)^2 } \,,
\end{equation}
which, compared to Eq.~\eqref{eq:def_D_L}, immediately implies that the luminosity distance in $D$-dimensions is given by
\begin{equation}
	d_L^{(D)} = a_0 r_N (1+z)^{2/(N-1)} = a_0 r_N (1+z)^{2/(D-2)} \,.
\end{equation}
This formula was first derived in \cite{1992ApJ...397....1C} for a general $D$-dimensional FRW metric including spatial curvature.
In 4-dimensions one correctly recovers the usual relation
\begin{equation}
	d_L^{(4)} = a_0r_3 (1+z) \,.
	\label{eq:dL_4}
\end{equation}

\bibliographystyle{JHEP}
\bibliography{draft}
\end{document}